\documentclass[prb,aps,twocolumn,showpacs, floatfix,preprintnumbers]{revtex4}

\usepackage[applemac]{inputenc}
\usepackage[T1]{fontenc} 
 \usepackage{amsmath} 
\usepackage{subfigure}
\usepackage{lipsum}
\usepackage{amsfonts} 
\usepackage{amssymb, mathrsfs}
\usepackage{braket}
\usepackage{graphicx} 
\usepackage[usenames]{color} 

\usepackage{bbm}

\DeclareMathOperator\sn{sn}
\DeclareMathOperator\cn{cn}
\DeclareMathOperator\dn{dn}

\DeclareMathOperator\arccot{arccot}
\def\beq{\begin{equation}}
\def\eeq{\end{equation}}
\def\bsp{\begin{split}}
\def\esp{\end{split}}
\def\bea{\begin{eqnarray}}
\def\eea{\end{eqnarray}}
\def\ba{\begin{array}}
\def\ea{\end{array}}

\def\lb{\left(}
\def\rb{\right)}

\def\l.{\left.}
\def\r.{\right.}

\def\part{\partial}

\def\bra#1{{\cal{h}} #1 \mid}
\def\ket#1{\mid #1 {\cal{i}}}

\def\Z{\ensuremath{\mathbb{Z}}}

\makeatletter

\newcommand{\Rmnum}[1]{\expandafter\@slowromancap\romannumeral #1@}
\makeatother
\begin{document}
\preprint{UdeM-GPP-TH-13-229}
\title{Quantum-classical transition of the escape rate of a biaxial ferromagnetic spin with an external magnetic field.}
\author{S. A. Owerre}
\email{solomon.akaraka.owerre@umontreal.ca}
\author{M. B. Paranjape} 
\email{paranj@lps.umontreal.ca}
\affiliation{Groupe de physique des particules, D\'epartement de physique,
Universit\'e de Montr\'eal,
C.P. 6128, succ. centre-ville, Montr\'eal, 
Qu\'ebec, Canada, H3C 3J7 }

\begin{abstract}
\section*{Abstract} 
We study the model of a biaxial single ferromagnetic spin Hamiltonian with an external magnetic field applied along the medium axis. The phase transition of the escape rate is investigated. Two different but equivalent methods are implemented. Firstly, we derive the semi-classical description of the model which yields a potential and a coordinate dependent mass. Secondly, we employ the method of spin-particle mapping which yields a similar potential to that of semi-classical description but with a constant mass. The exact instanton trajectory and its corresponding action, which have not been reported in any literature is being derived. Also, the analytical expressions for the first- and second-order crossover temperatures at the phase boundary are derived.  We show that the boundary between the first-and the second-order phase transitions is greatly influenced by the magnetic field. 

\end{abstract}


\maketitle


{$\mathbf{Introduction}$-} In recent years, the study of single ferromagnetic spin systems has been  of considerable interest to condensed matter physicists. These systems  have been pointed out\cite{chud2,chud1} to be a good candidate for investigating  first- and second-order phase transition of the quantum-classical escape rate. The quantum-classical escape rate transition takes place in the presence of a potential barrier. At very low temperature (close to zero), transitions occur by quantum tunnelling through the barrier and the rate is governed by $\Gamma\sim e^{-B}$, where $B$ is the instanton (imaginary time solution of the classical equation of motion) action. At high temperatures, the particle has the possibility of hopping over the barrier (classical thermal activation), in this case transition is governed by $\Gamma\sim e^{-\frac{\Delta V}{T}}$ , where $\Delta V$ is the energy barrier. At the critical point when these two transition rates are equal, there exits a crossover temperature (first-order transition) $T_0^{(1)}$ from quantum to thermal regime, it is estimated as $T_0^{(1)}=\Delta V/B$. In principle these transitions are greatly influenced by the anisotropy constants and the external magnetic fields. The second-order phase transition occurs for  particles in a cubic or quartic parabolic potential, it take place at the temperature $T_{0}^{(2)}$, below $T_{0}^{(2)}$  one has the phenomenon of thermally assisted tunnelling and above $T_{0}^{(2)}$ transition occur due to thermal activation to the top of the potential barrier\cite{ chud2,chud1}. The order of these transitions can also be determined from the period of oscillation $\tau(E)$ near the bottom of the inverted potential. Monotonically increasing $\tau(E)$ with the amplitude of oscillation gives a second-order transition while nonmonotonic behaviour of $\tau(E)$ ( that is a mininmum in the $\tau(E)$ vs $E$ curve, $E$ being the energy of the particle ) gives a first-order transition\cite{chud2}. 

The model of a uniaxial single ferromagnetic spin with a transverse magnetic field, which is believed to describe the molecualr magnet MnAc$_{12}$ was considered by Garanin and Chudnovsky\cite{chud2}, the Hamiltonian is of the form $\hat{H}= -D\mathcal{\hat{S}}_{z}^2 - h_x\mathcal{\hat{S}}_x$,  using the spin-particle mapping version of this Hamiltonian \cite{solo,solo1, wznw}, they showed that the transition from thermal to quantum regime is of first-order in the regime $h_x<sD/2$ and of second-order in the regime $sD/2 < h_x<2sD$.  For other single-molecule magnets such as Fe$_{8}$, a biaxial ferromagnetic spin model is a good approximation. In this case, Lee $\textit{et al}$\cite{c} considered the model $\hat{H}= K(\mathcal{\hat{S}}_{z}^2 +\lambda\mathcal{\hat{S}}_{y}^2) - 2\mu_Bh_y\mathcal{\hat{S}}_y$, using spin coherent state path integral, they obtained a potential and a coordinate dependent mass from which they showed that the boundary between the first and the second-order transitions sets in at $\lambda=0.5$ for $h_y=0$ while the order of the transitions is greatly influenced by the magnetic field and the anisotropy constants for $h_y\neq0$.  Zhang $\textit{et al}$\cite{solo6} studied the model  $\hat{H}= K_1\mathcal{\hat{S}}_{z}^2 +K_2\mathcal{\hat{S}}_{y}^2 $ using spin-particle mapping and periodic instanton method. The phase boundary between the first- and the second-order transitions was shown to occur at $K_2=0.5K_1$. The model with $z$-easy axis in an applied field has been also studied by numerical and perturbative methods\cite{eg}. In this paper, we study  a biaxial spin system with an external magnetic field applied along the  medium axis using spin-coherent state path integral and the  formalism of spin-particle mapping. Unlike other models with an external magnetic field\cite{solo4,solo6,kal}, the spin-particle mapping yields a simplified potential and a constant mass which allows us to solve for the exact instanton trajectory and its corresponding action in the presence of a magnetic field. We also present the analytical results of the crossover temperatures for the first- and the second-order transitions at the phase boundary.

{$\mathbf{Spin \thinspace model \thinspace and  \thinspace spin\thinspace coherent \thinspace state \thinspace path \thinspace integral}$-}
Consider the Hamiltonian of a biaxial ferromagnetic spin (single-molecule magnet) in an external magnetic field 
\begin{align}
\hat{H}&= \mathcal{D}\mathcal{\hat{S}}_{z}^2  + \mathcal{E}\mathcal{\hat{S}}_{x}^2  - h_x\mathcal{\hat{S}}_x  \label{1}\end{align}
where $\mathcal{D}\gg\mathcal{E}>0$, and $\mathcal{S}_i, i=x,y,z$ is the components of the spin. This model possesses an easy $XOY$ plane with an easy-axis along the $y$-direction and an external magnetic field along the $x$-axis.  At zero magnetic field, there are two classical degenerate ground states corresponding to the minima of the energy located at $\pm y$, these ground states remain degenerate for $h_x\neq0$ in the easy $XY$ plane. The semi-classical form of the quantum Hamiltonian can be derived using spin coherent state path integral. In the coordinate dependent form, the spin-coherent-state is defined by \cite{ams,ams1}
\bea
 \ket{\bold{\hat{n} }} = \left(\cos\frac{1}{2}\theta \right)^{2s}\exp\left\lbrace\tan\left(\frac{1}{2}\theta \right)e^{i\phi }\mathcal{\hat{S}} ^{-}\right\rbrace\ket{s ,s }
 \eea
where $\bold{\hat{n}}  = s(\sin\theta \cos\phi ,\sin\theta\sin\phi ,\cos\theta )$ is the unit vector parametrizing the spin on a two-sphere $S^2$. The overlap between two coherent states is found to be
\bea \braket{\bold{\hat{n}}^{\prime}|\bold{\hat{n}}} =\Bigg[\cos\frac{1}{2}\theta \cos\frac{1}{2}\theta ^{\prime} +\sin\frac{1}{2}\theta \sin\frac{1}{2}\theta ^{\prime}e^{-i\Delta \phi }\Bigg]^{2s}
\label{2.4}
\eea
where $\Delta \phi  = \phi ^{\prime}  -\phi$. The expectation value of the spin operator in the large $s$ limit is approximated as
$
\braket{\bold{\hat{n}}^{\prime}|\mathcal{\hat{\boldsymbol{S}}}|\bold{\hat{n}}} \approx s\left[\bold{\hat{n}} +O\left(\sqrt{s} \right)\right]\braket{\bold{\hat{n}}^{\prime}|\bold{\hat{n}}} 
$
. For infinitesimal separated angle, $\Delta \theta = \theta ^{\prime} -\theta$, Eq.\eqref{2.4} reduces to 

\bea \braket{\bold{\hat{n} }^{\prime}|\bold{\hat{n}} }\approx 1 -i s \Delta\phi (1-\cos\theta ).\eea

These states satisfy the overcompletness relation (resolution of identity)
\begin{align}
\mathcal{N}\int   \thinspace d\phi  \thinspace d(\cos \theta )\ket{\bold{\hat{n}} }\bra{\bold{\hat{n}} }  =\hat{I}.
\end{align}
Using these equations, the transition amplitude is easily obtained as
\bea \braket{\bold{\hat{n}}_{ f}|e^{- \beta \hat{H} }|\bold{\hat{n}}_{ i}} =   \int\mathscr{D}\phi \mathscr{D}(\cos\theta )e^{-S}
\label{3.5a}
\eea

The Euclidean action $(t\rightarrow-i\tau)$  is 
given by $S =  \int_{-\beta/2}^{\beta/2}d\tau\thinspace \mathcal{L}, $ with
\bea
\mathcal{L}= is\dot{\phi} (1 -\cos\theta )  + V\left(\theta , \phi  \right)
\label{3.6}
\eea

\begin{align}
V\left(\theta , \phi  \right)&= \mathcal{D} s^2 \cos^2 \theta  +\mathcal{E}s^2\sin^2\theta\cos^2\phi-sh_x\sin\theta\cos\phi
\label{3.6a}
\end{align}

These two equations \eqref{3.6} and \eqref{3.6a} describe the semi-classical dynamics of the spin on $S^2$. Two degenerate minima exit for $ h_x<h_c=2\mathcal{E}s$, which are located at $\theta= \pi/2$: $\phi=2\pi n \pm\arccos\alpha_x$, where $\alpha_x= h_x/h_c$, $n\in \Z$,    and the of the maximum is at  $\theta= \pi/2$: $\phi=n\pi$ with the height of the barrier ($n=0$) given by
\bea
\Delta V = \mathcal{E}s^2(1-\alpha_x)^2
\eea
Taking into consideration the fact that $\mathcal{D}\gg \mathcal{E}$, the deviation away from the easy plane is very small, thus one can expand $\theta= \pi/2-\eta$, where $\eta\ll 1$. Integration over the fluctuation $\eta$ in Eq.\eqref{3.5a} yields an effective theory describe by
\bea
\mathcal{L}_{\text{eff}}=is\dot{\phi}+ \frac{1}{2}m(\phi)\dot{\phi}^2+ V(\phi)
\eea
where
\bea
V(\phi) = \mathcal{E}s^2(\cos\phi-\alpha_x)^2
\label{pot1}
\eea
and
\bea
m(\phi) = \frac{1}{2\mathcal{D}(1-\kappa\cos^2\phi+2\alpha_x\kappa\cos\phi)}
\label{mas1}
\eea
with $\kappa=\mathcal{E}/\mathcal{D}$. An additional constant  of the form $\mathcal{E}s^2\alpha_x^2$ has been added to the potential for convenience. The first term in the effective Lagrangian is a total derivative which does not contribute to the classical equation of motion, however, it has a significant effect in the quantum transition amplitude, producing a quantum phase interference in spin systems \cite{cl,l}. The two classical degenerate minima which corresponds to $\phi= 2\pi n \pm \arccos\alpha_x$  are separated by a small barrier at $\phi=0$ and a large barrier at $\phi=\pi$. The phase transition of the escape rate of this model can be investigated using the potential Eq.\eqref{pot1} and the mass Eq.\eqref{mas1} \cite{c}, in this paper, however, we will study this transition via the method of mapping a spin system onto a quantum mechanical particle in a potential field.  A classical trajectory (instanton) exits for zero magnetic field, in this case the classical equation of motion
\bea
m(\bar{\phi})\ddot{\bar{\phi}}+ \frac{1}{2}m(\bar{\phi})^{\prime}\dot{\bar{\phi}}= \frac{dV}{d\bar{\phi}}
\label{cla}
\eea
integrates to 
\bea
\sin\bar{\phi} = \pm \frac{\sqrt{(1-\kappa)}\tanh(\omega\tau)}{\sqrt{1-\kappa\tanh^2(\omega\tau)}}
\label{int1}
\eea
where $\omega =  2s\sqrt{\mathcal{E}\mathcal{D}}$ and the upper and lower signs are for instanton and anti-instanton respectively. The corresponding action for this trajectory yields\cite{chud,cl} $S_0= B\pm is\pi$, 
\bea
B= s\ln\lb\frac{1+\sqrt{\kappa}}{1-\sqrt{\kappa}}\rb
\label{2.15}
\eea
For small anisotropy parameters, $\kappa\ll 1$,  the coordinate dependent mass can be approximated as $m\approx 1/2\mathcal{D}$, the approximate instanton trajectory in this limit yields
\bea
\sin\bar{\phi}=\pm \frac{2 \sqrt{\frac{1-\alpha_x}{1+\alpha_x}}\tanh(\omega\tau)}{[1+\frac{1-\alpha_x}{1+\alpha_x}\tanh^2(\omega\tau)]}
\label{2.16}
\eea

where $\omega=s\sqrt{\mathcal{E}\mathcal{D}(1-\alpha_x^2)}$ and the corresponding action is
\begin{align}
B= 2s\sqrt{\kappa}[\sqrt{1-\alpha_x^2}\pm\alpha_x \arcsin(\sqrt{1-\alpha_x^2})] 
\label{2.17}
\end{align}
 The upper and the lower signs in the action correspond to the large and small barriers respectively while that in the trajectory is for  instanton and anti-instanton. 
 At zero magnetic field, the instanton interpolates between the classical degenerate minima $\bar{\phi}=\pm\pi/2$ at $\tau=\pm\infty$. For coordinate dependent mass the classical trajectory can be integrated in terms of the Jacobi elliptic functions. This solution will be presented  in  the next section using a simpler method.
 
{$\mathbf{Particle \thinspace mapping}$ -}
In this section, we will consider the formalism of mapping a spin system to a quantum-mechanical particle in a  potential field \cite{solo}. In this formalism one introduces a nonnormalized spin coherent state, the action of the spin operators on this state yields the following expressions\cite{solo1,wznw}
\begin{align}
&\mathcal{\hat{S}}_x = s\cos\phi-\sin\phi\frac{d}{d\phi}, \thinspace \mathcal{\hat{S}}_y = s\sin\phi+\cos\phi\frac{d}{d\phi}\nonumber\\&\mathcal{\hat{S}}_z =  -i\frac{d}{d\phi}
\label{4.1}
\end{align}
  The Shr\"{o}dinger equation can be written as
  \bea
  \hat{H}\Phi(\phi) = E\Phi(\phi)
  \label{4.2}
  \eea
  where the generating function is defined as  
\begin{align}
\Phi(\phi) =\sum_{m=-s}^{s}\frac{\mathcal{C}_{m}}{\sqrt{(s-m)!(s+m)!}}e^{im\phi} 
  \label{2.6}
\end{align}
 with periodic boundary condition $\Phi(\phi+2\pi)$= $e^{2i\pi s}\Phi(\phi)$. Using Eqns.\eqref{1}, \eqref{4.1} and \eqref{4.2},  the differential equation for $\Phi(\phi)$ yields
\begin{equation}
\begin{split}
& -\mathcal{D}(1+\kappa\sin^2\phi)\frac{d^2\Phi}{d\phi}-(\mathcal{E}(s-\frac{1}{2})\sin 2\phi-h_x\sin\phi )\frac{d\Phi}{d\phi}\\&+(\mathcal{E}s^2\cos^2\phi +\mathcal{E}s\sin^2\phi-h_xs\cos\phi )\Phi= E\Phi
\label{3.4b} 
\end{split}
\end{equation} 
Now let's introduce the incomplete elliptic integral of first kind 
 \beq
 x = F(\phi,\lambda)= \int_{0}^{\phi}d\varphi  \frac{1}{\sqrt{1-\lambda^2\sin^2\varphi}}
  \label{3.15}
  \eeq
 with amplitude $\phi$ and modulus $\lambda^2=\kappa$. The trigonometric functions are related to the Jacobi elliptic functions by $\sn(x,\lambda)=\sin\phi$, $\cn(x,\lambda)=\cos\phi$ and $\dn(x,\lambda)=\sqrt{1-\lambda^2\sn^2(x,\lambda)}$.
In this new variable, Eq.\eqref{3.4b} transforms into a  Schr\"{o}dinger equation $H\Psi(x)=E \Psi(x)$
with
\beq
H = - \frac{1}{2m}\frac{d^2}{dx^2}  + V(x) , \quad m= \frac{1}{2\mathcal{D} } 
\label{3.18}
\eeq
The effective potential is given by
\bea
V(x) = \frac{\mathcal{E}\tilde{s}^2[\cn(x,\lambda)-\alpha_x]^2}{\dn^2(x,\lambda)}
\label{3.1}
\eea
 
\begin{align}
& \Psi(x)=\frac{\Phi(\phi(x))}{ [\dn(x,\lambda)]^{s }} \exp\bigg[-\tilde{s}\alpha_x\sqrt{\frac{\kappa}{(1-\kappa)}}\nonumber\\&\arccot\lb \sqrt{\frac{\kappa}{(1-\kappa)}}\cn(x,\lambda)\rb\bigg]
\end{align}
where $\tilde{s}=(s+\frac{1}{2})$ and $\alpha_x = h_x/2\mathcal{E}\tilde{s}$. In order to arrive at this potential we have used the large $s$ limit $s(s+1)\sim \tilde{s}^2$ and shifted the minimum energy to zero by  adding a constant  of the form $\mathcal{E}\tilde{s}^2\alpha_x^2$.
Unlike the spin coherent state version, the mass of the particle is constant in this case which appears to be the approximate form of Eq.\eqref{mas1} in the limit of small anisotropy parameters, but the potentials Eq.\eqref{pot1} and Eq.\eqref{3.1} are of similar form, infact they are equal when $\lambda\rightarrow0$ except for the quantum renormalization $\tilde{s}$.  At zero magnetic field the potential  Eq.\eqref{3.1} reduces  to a well-known potential studied by periodic instanton method \cite{solo6}. In many models with an external magnetic field\cite{solo4,solo6,kal}, the resulting effective potential from spin-particle mapping is always too complicated for one to solve for the instanton trajectory, however in this case the effective potential is in a compact form, allowing us to find the exact classical trajectory (see the next section).

{$\mathbf{Phase \thinspace transition \thinspace and   \thinspace instanton\thinspace solution}$-}
We will now study the phase transition of the escape rate of this model and the instanton solution in the presence of a magnetic field. The potential Eq.\eqref{3.1} has minima at $x_0=4n\mathcal{K(\lambda)}\pm\cn^{-1}(\alpha_x)$ and maxima at $x_{sb}=\pm4n\mathcal{K(\lambda)}$ for small barrier and at $x_{lb}=\pm2(2n+1)\mathcal{K(\lambda)}$ for large barrier, where $\mathcal{K(\lambda)}$  is the complete elliptic function of first kind i.e $F(\frac{\pi}{2},\lambda)$. The heights of the potential for small and large barriers are given by
\begin{align}
\Delta V_{sb} = \mathcal{E}\tilde{s}^2(1-\alpha_x)^2\nonumber\\
\Delta V_{lb} = \mathcal{E}\tilde{s}^2(1+\alpha_x)^2
\label{ba}
\end{align}
\begin{figure}[ht]
\centering
\includegraphics[scale=0.35]{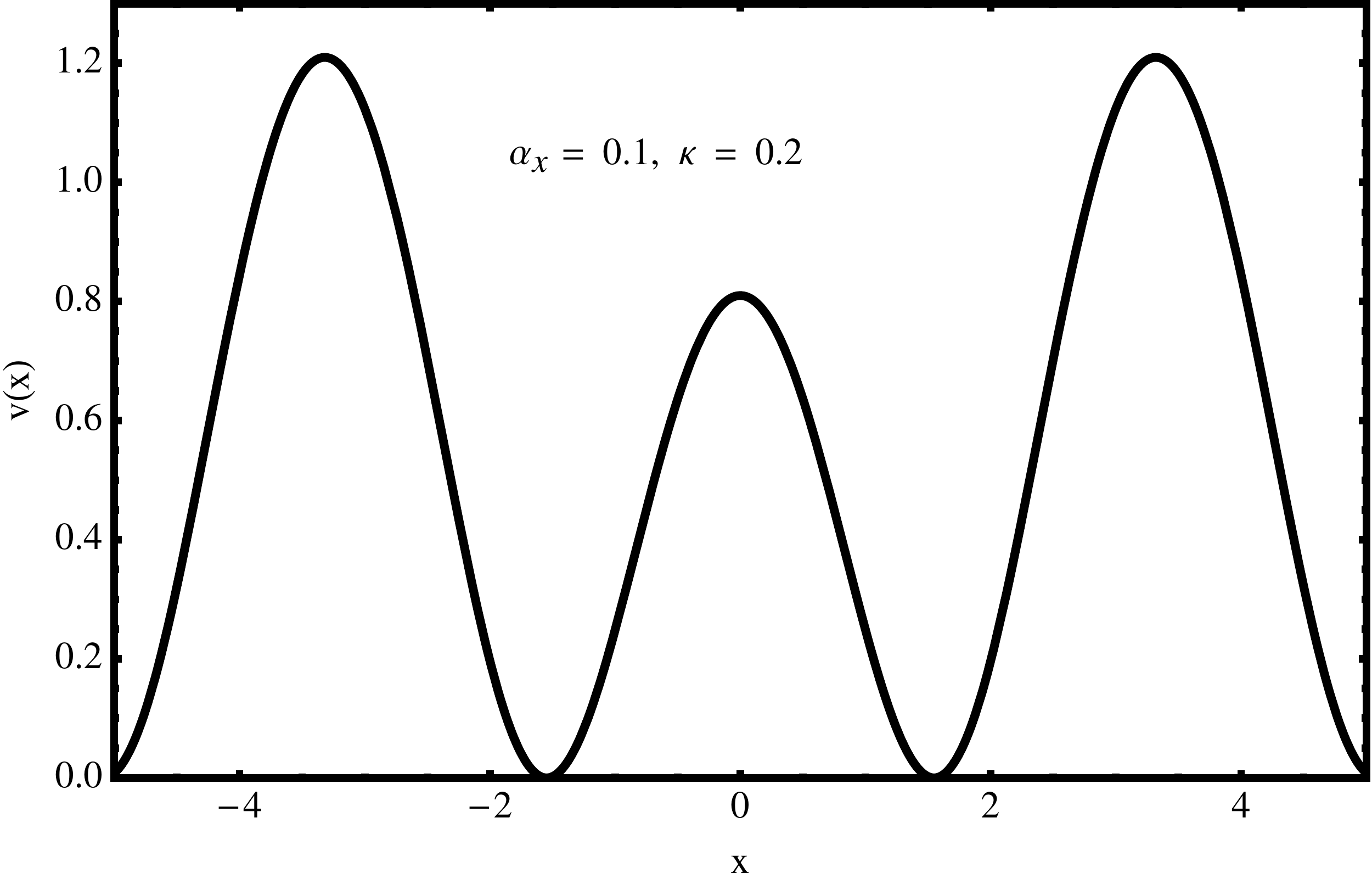}
\label{pont}
\caption{ The plot of the effective potential, Eq.\eqref{3.1} for $\alpha_x=0.1$, $\kappa=0.2$, where $v(x)= V(x)/\mathcal{E}\tilde{s}^2$.}
\end{figure}
The Euclidean Lagrangian corresponding to the particle Hamiltonian is
 
\bea
\mathcal{L} = \frac{1}{2}m\dot{x}^2 + V(x)
\eea
It follows that the classical equation of motion is
\bea
m\ddot{\bar{x}}=\frac{dV}{d\bar{x}}
\label{3.4}
\eea
which corresponds to the motion of the particle in the inverted potential $-V(x)$. Upon integration, Eq.\eqref{3.4} gives the instanton solution
\bea
 \sn(\bar{x},\lambda) =\pm \frac{2 \sqrt{\frac{1-\alpha_x}{1+\alpha_x}}\tanh(\omega\tau)}{[1+\frac{1-\alpha_x}{1+\alpha_x}\tanh^2(\omega\tau)]} 
\label{int}
\eea
where $\omega= \tilde{s}\sqrt{\mathcal{E}\mathcal{D} (1-\alpha_x^2)}$. 
This trajectory has not been reported in any literature. It is the exact classical trajectory in the presence of an external magnetic field. The instanton (upper sign) interpolates from the left minimum $\bar{x}(\tau)=-\sn^{-1}(\sqrt{1-\alpha_x^2})$ at $\tau=-\infty$ to the center of the barrier $\bar{x}(\tau)=0$ at $\tau=0$ and reaches the right minimum $\bar{x}(\tau)=\sn^{-1}(\sqrt{1-\alpha_x^2})$ at $\tau=\infty$. At zero magnetic field, Eq.\eqref{int} is equivalent to the well-known instanton solution \cite{solo2}, which is equivalent to Eq.\eqref{int1}. It is noted that this trajectory is the same as Eqn.\eqref{2.16} except that the trigonometric sine function is being replaced by the Jacobi elliptic sine function and $s\rightarrow \tilde{s}$, however,  in the limit $\lambda\rightarrow0$, both solutions are the same, since the potentials Eqns.\eqref{pot1} and \eqref{3.1} and the masses  Eqn.\eqref{mas1} and Eqn.\eqref{3.18} are the same in this limit (the Jacobi elliptic functions becomes the trigonometric functions). The action for the trajectory, Eq.\eqref{int} yields
\begin{align}
B&= \tilde{s}\bigg[ \ln\lb\frac{1+\sqrt{\kappa(1-\alpha_x^2)}}{1-\sqrt{\kappa(1-\alpha_x^2)}}\rb\label{act}\nonumber\\&\pm 2\alpha_x \sqrt{\frac{\kappa}{1-\kappa}}\arctan\lb \frac{\sqrt{(1-\kappa)(1-\alpha_x^2)}}{\alpha_x}\rb\bigg]
\end{align}
 When $\alpha_x=\pm1$, there is no large and small barriers, the trajectory and its action reduce to  $\bar{x}(\tau)=0=B$, hence there is no tunnelling.   It is noted that this action reduces to Eq.\eqref{2.17}  in the limit $\kappa\ll 1$ and to Eq.\eqref{2.15} when $\alpha_x= 0$ except that $s$ is being replaced by $\tilde{s}$.
 At nonzero energy (finite temperature), the particle has the possibility of hopping over the potential barrier (thermal activation), the escape rate (transition amplitude) of the particle can be either first- or second-order depending on the shape of the potential. In order to investigate the  analogy of this transition to Landau's theory of phase transition, consider the     the escape rate of a particle at finite temperature through a potential barrier in the quasiclassical approximation \cite{cl,chud1}
\begin{align}
\Gamma \sim \int dE \thinspace \mathcal{W}(E) e^{-(E-E_{\text{min}})/T}
\end{align}
where $\mathcal{W}(E)$ is the tunnelling probability of a particle at an energy $E$, and $E_{\text{min}}$ is the energy at bottom of the potential.  The tunnelling probability in imaginary time is given  as
$
\mathcal{W}(E) \sim e^{-S(E)}, 
$
therefore we have
\begin{align}
\Gamma \sim e^{-F_{\text{min}}/T}
\end{align}
where $F_{\text{min}}$ is the minimum of the free energy $F \equiv E + TS(E)- E_{\text{min}}$ with respect to $E$. The imaginary time action is expressed as
\bea
S(E)=2\sqrt{2m}\int_{-x(E)}^{x(E)} dx \sqrt{V(x)-E}
\eea
where $\pm x(E)$ are the turning point for the particle with energy $-E$ in an inverted potential. Introducing a dimensionless quantity $Q= (V_{\text{max}}-E)/(V_{\text{max}}-V_{\text{min}})$ where $V_{\text{max}}(V_{\text{min}})$ corresponds to the top (bottom) of the potential, the expansion of the imaginary time action around $x_b$ gives \cite{solo4} 
\bea
S(E)= \frac{2\pi\Delta V}{\omega_0}[Q^2 + b Q^2+ O(Q^3)]
\label{ac}
\eea
where
\begin{align}
b&= \frac{\Delta V}{48(V^{\prime\prime}(x))^3}[  5(V^{\prime\prime\prime }(x))^2-3V^{\prime\prime\prime\prime}(x) V^{\prime\prime}(x) ]_{x=x_b}\nonumber
\end{align}
and $\omega_0^2=-V^{\prime\prime}(x_b)/{m}>0$ is the frequency of oscillation at the bottom of the inverted potential, $x_b$ corresponds to the maximum of the potential. 
\begin{figure}[ht]
\centering
\subfigure[ ]{%
\includegraphics[scale=0.35]{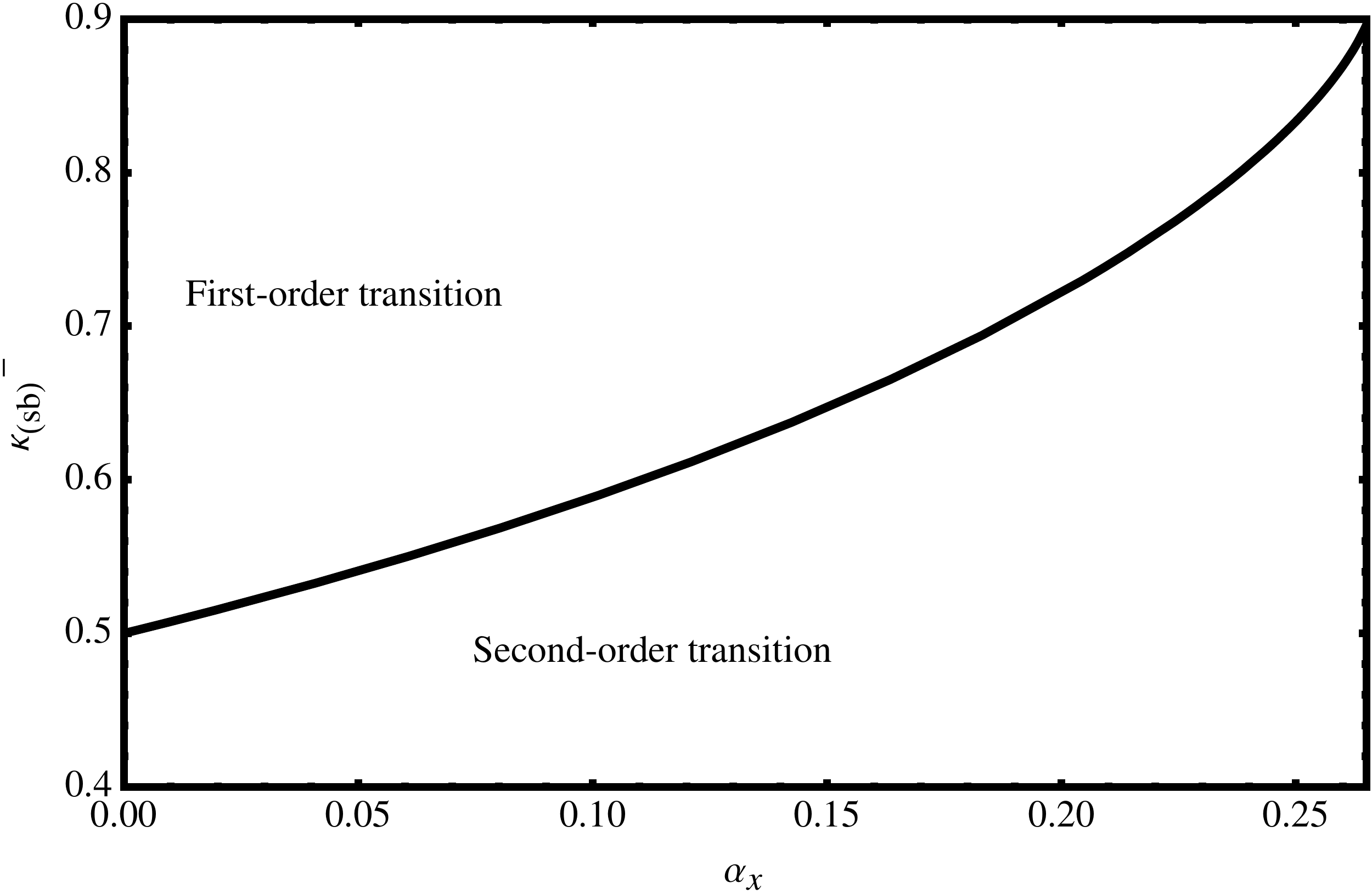}
\label{fr}}
\subfigure[]{%
\includegraphics[scale=0.35]{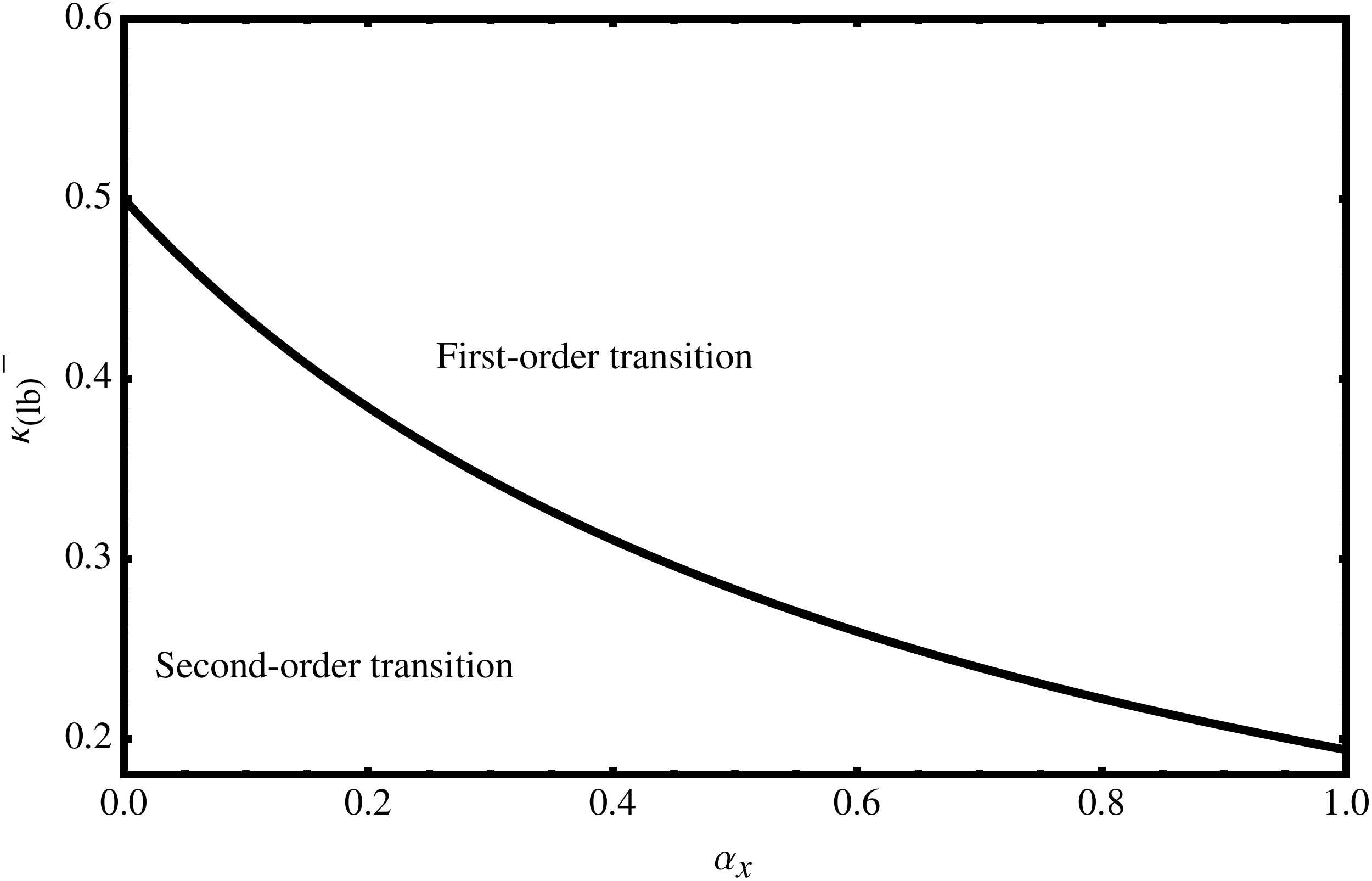}
\label{phase}}
\caption{  The phase diagram  $\kappa^{-}$ vs $\alpha_x$ at the phase boundary for small barrier (a) and large barrier (b).}
\label{ph}
\end{figure}

By the analogy with the Landau theory of phase transition, the phase boundary between the first- and second-order transition (see Fig.(1)) is obtained by setting the coefficient  of $Q^2$ to zero i.e $b=0$. Using the maximum of the small and large barriers of the potential Eq.\eqref{3.1} at $x_{sb}$ and $x_{lb}$ we obtain
\begin{align}
b_{sb}= (\kappa-\kappa_{sb}^{+}(\alpha_x))(\kappa-\kappa_{sb}^{-}(\alpha_x))\\
b_{lb}= (\kappa-\kappa_{lb}^{+}(\alpha_x))(\kappa-\kappa_{lb}^{-}(\alpha_x))
\end{align}
where
\begin{align}
\kappa_{sb}^{\pm}(\alpha_x)=\frac{3 - 4 \alpha_x +  \alpha_x^2 \pm (1-\alpha_x)\sqrt{1 - 4\alpha_x +  \alpha_x^2 }}{4(1- 2 \alpha_x +  \alpha_x^2)}\label{3.9a}
\\
\kappa_{lb}^{\pm}(\alpha_x)=\frac{3 + 4 \alpha_x +  \alpha_x^2 \pm (1+\alpha_x)\sqrt{1 + 4\alpha_x +  \alpha_x^2 }}{4(1+2 \alpha_x +  \alpha_x^2)}\label{3.9b}
\end{align}
Thus by setting $b=0$ we obtain the four solution in Eqns.\eqref{3.9a} and\eqref{3.9b}. At $\alpha_x=0$, the critical values at the phase boundary are $\kappa_c=$ $1$ or $\frac{1}{2}$ for the plus or the minus signs respectively\cite{l1,solo6,solo4}. Expanding for small field $\alpha_x\ll1$, we obtain $\kappa_{sb/lb} ^{+} \approx 1\pm\frac{\alpha_x}{4}$ and $\kappa^{-}_{sb/lb} \approx \frac{1}{2}(1\pm\frac{3}{2}\alpha_x)$, where the plus and minus signs correspond to the small and large barriers respectively.
\begin{figure}[ht]
\centering
\subfigure[ ]{%
\includegraphics[scale=0.35]{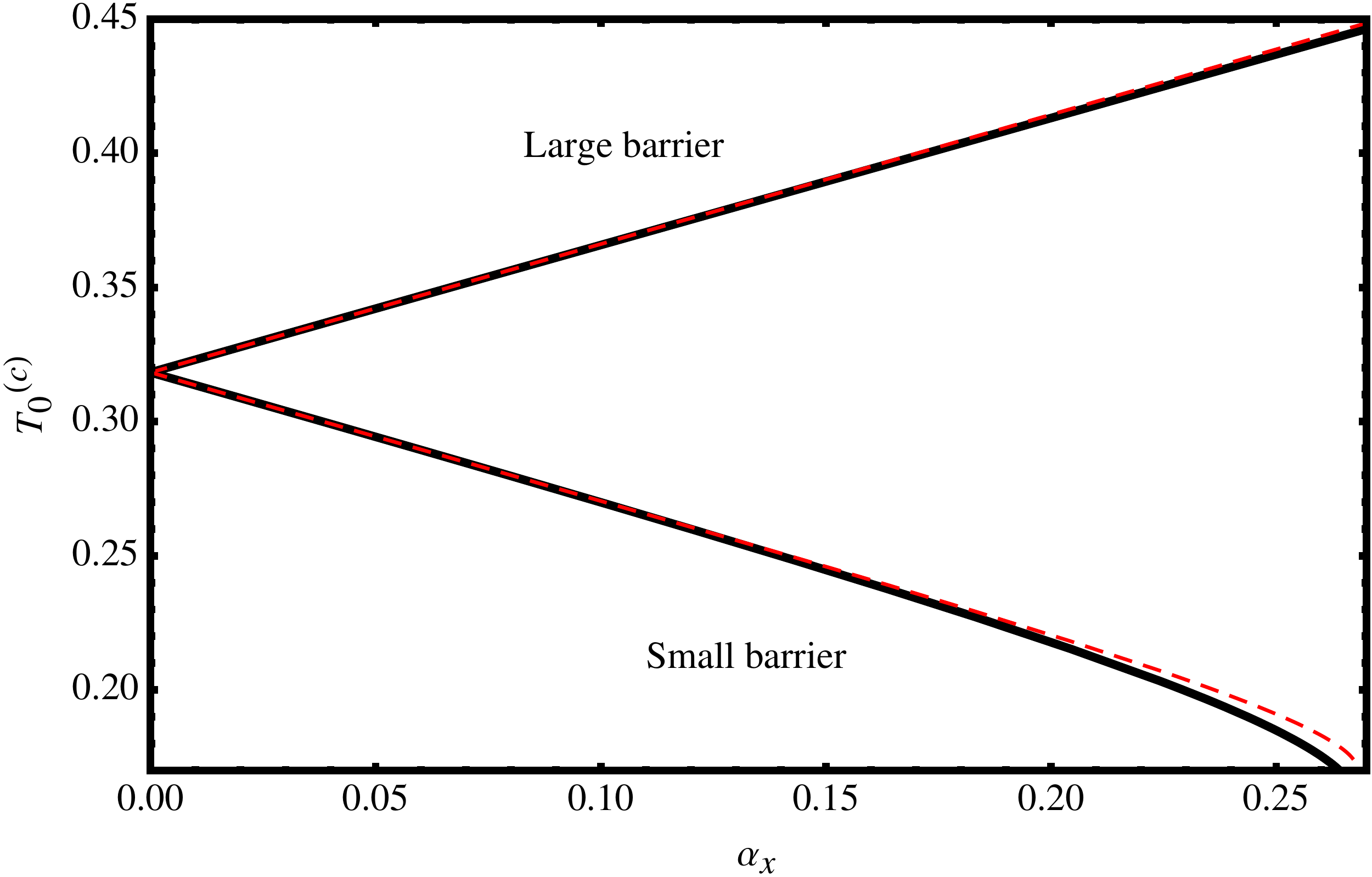}
\label{phase0}}
\subfigure[]{%
\includegraphics[scale=0.35]{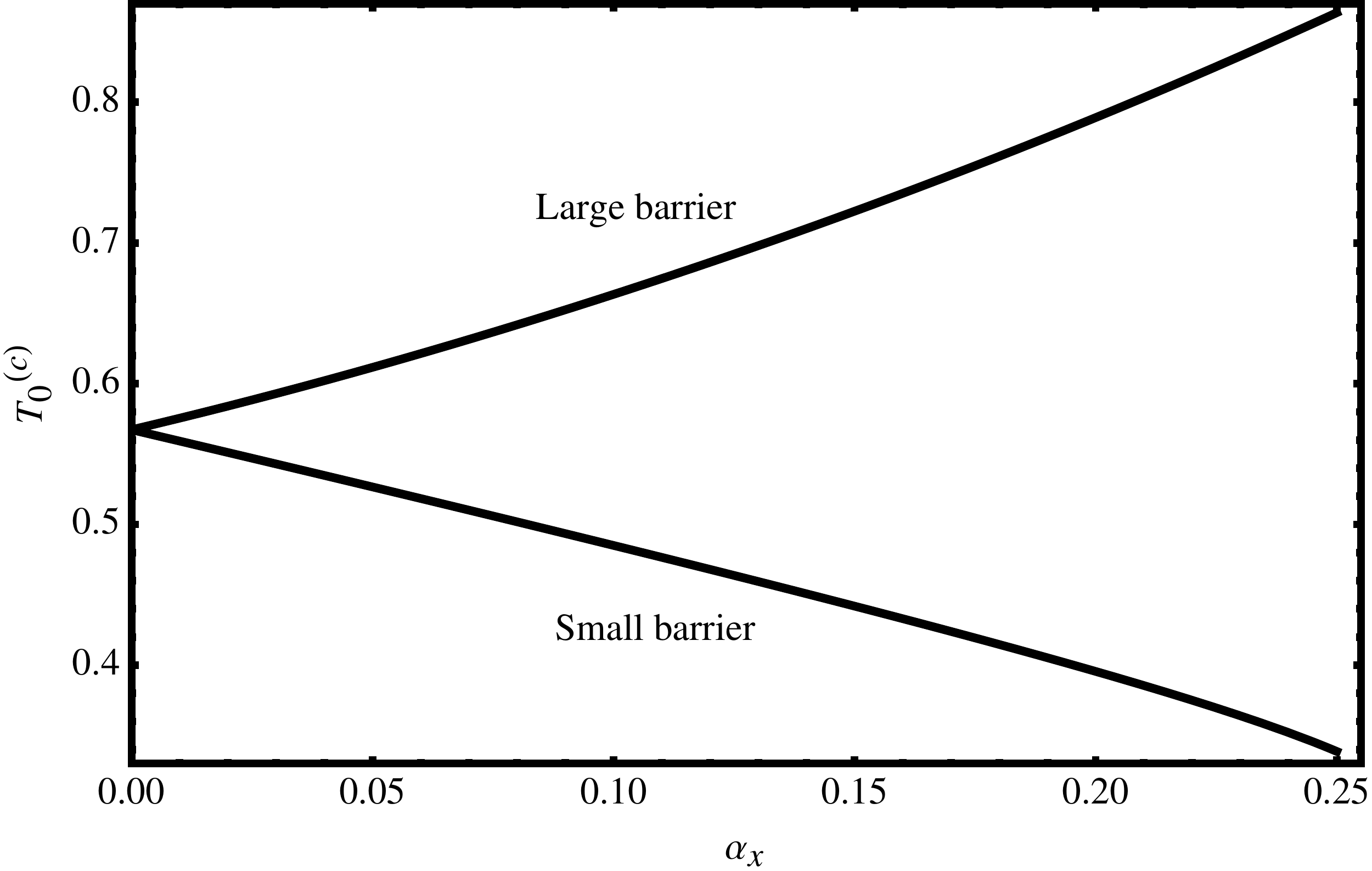}
\label{phase1}}
\caption{ Colour online: Dependence of the crossover temperatures on the magnetic field at the phase boundary:  (a) Second-order  (solid line) and its maximum (dashed line) for the small and large barrier, (b) First-order for the small and the large barrier. These graphs are plotted with $T_0^{(c)}= T_0^{(c)}/ \mathcal{E}\tilde{s}$.}
\label{phase}
\end{figure}
The phase diagrams of Eqns.\eqref{3.9a}  and\eqref{3.9b} are shown in Fig.\eqref{ph}, with the value $\kappa^{-}$ increasing with increasing magnetic field for small barrier while it decreases  with increasing magnetic field for large barrier, the first-order phase transition occurs in the regime $\kappa^{-}_{sb/lb}>1/2$ in both cases. The crossover temperature for the first-order transition is estimated as
$
T_0^{(1)}= \Delta V/B
$
which is easily obtained from Eqns.\eqref{ba} and
\eqref{act}. Expanding for $\alpha_x\ll1$ at the phase boundary (with the expressions for $\kappa^{-}_{sb/lb}(\alpha_x)$), we obtain the crossover temperatures as $T_0^{(c)}\approx\mathcal{E}\tilde{s}/(\ln[(3+2\sqrt{2})e^{\pm \frac{3\alpha_x}{\sqrt{2}}}])$, where the upper and lower signs correspond to   small and large barrier respectively. Both temperatures coincide at $ \alpha_x=0\Rightarrow\kappa^{-}_{sb/lb}=1/2$ with $T_0^{(c)}=\mathcal{E}\tilde{s}/\ln(3+2\sqrt{2})$   as shown in Fig.\eqref{phase1}. 
In the case of second-order transition the crossover temperature is estimated as  $  T_{0}^{(2)}= \omega_0/{2\pi}$. This is easily obtained as

\bea
  T_{0}^{(2)}=  \frac{\mathcal{E}\tilde{s}\sqrt{ (1\pm \alpha_x)}}{ \pi}\lb\frac{1-\lb 1\pm\alpha_x\rb\kappa }{\kappa }\rb^{1/2}
  \label{sec}
\eea

The maximum of this function occurs at $\alpha_x= \pm(1-2\kappa)/2\kappa$, with 
\bea
  T_{0}^{(\text{max})}=  \frac{\mathcal{E}\tilde{s}}{ 2\pi\kappa} 
  \label{max}
  \eea
  where the upper and lower signs correspond to the large and small barriers respectively.
 Substituting the expressions for $\kappa^{-}_{sb/lb}(\alpha_x)$ into Eqns. \eqref{sec} and \eqref{max} we obtain the temperatures at the phase boundary as shown in 
Fig.\eqref{phase0}. The critical temperature at the phase boundary decreases with increasing magnetic field for small barrier while for large barrier it increases with increasing magnetic field.  In the regime of small field $\alpha_x\ll 1$, it behaves linearly as $T_{0}^{(c)}\approx \mathcal{E}\tilde{s}(1\pm\frac{3}{2}\alpha_x)/{\pi}$. Both barriers coincide at $ \alpha_x=0\Rightarrow\kappa^{-}_{sb/lb}=1/2$, with $T_{0}^{(c)}= \mathcal{E}\tilde{s}/{\pi}$ which is smaller than that of first-order.
 
 {$\mathbf{ Conclusions}$-}
In conclusion, we have investigated an effective particle Hamiltonian which corresponds exactly to a biaxial spin model. Using this Hamiltonian we studied the phase transition of the escape rate of a particle at zero and nonzero temperatures. The analytical expressions for the instanton trajectories and the crossover temperatures were obtained. We showed that the boundary between the first-and second-order phase transition is greatly influenced by the magnetic field.

{$\mathbf{ Acknowledgments}$-}
The authors would like to thank  NSERC of Canada for financial support. 


\vfill
 
\begin{thebibliography}{99}
 \bibitem{chud2} 
E.M. Chudnovsky and D.A. Garanin, \prl {\bf 79}, 4469 (1997).
\bibitem{eg} 
E.M. Chudnovsky and D.A. Garanin, \prb {\bf 59}, 3671 (1999); \prb {\bf 63}, 024418 (2000)
 \bibitem{chud1}
 D. A. Garanin, X. Mart\`inez Hidalgo, and E. M. Chudnovsky  \prb {\bf  57}, 13639 (1998)
\bibitem{solo4}
Gwang-Hee Kim, \prb {\bf 59},  11847, (1999); J. Appl. Phys. 86, 1062 (1999) 
\bibitem{solo}
G Scharf, W F Wreszinski and -J L van Hemmen, J. Phys. A: Math. Gen {\bf 20}, 4309 (1987) 

\bibitem{solo1}
O.B. Zaslavskii, Phys. Lett. A {\bf 145}, 471 (1990) 

\bibitem{wznw}
V.V. Ulyanov, O.B. Zaslavskii, Phys. Rep. {\bf 214}, 179 (1992)
\bibitem{solo2}
J.-Q. Liang, H. J. W. M\"{u}ller-Kirsten, D. K. Park and F.-C. Pu , \prb {\bf 61}, 8856 (2000) 
\bibitem{l1}
  Our model for $h_x=0$ is equivalent to that of  Ref. [14] if we set $K_1=\mathcal{D} $ and $K_2=\mathcal{E}$ and  it is equivalent to that of  Ref. [4] if we set $K_{\perp}=\mathcal{D}-\mathcal{E}$ and $K_{||}=\mathcal{E}$ .
  \bibitem{cl}
Daniel Loss, David P. DiVincenzo, and G. Grinstein, \prl {\bf 69}, 3232 (1992)
   
 \bibitem{l}
 J. von Delft, C. Henley, \prl {\bf 69}, 3236(1992)
\bibitem{kal}
Chang-Soo Park, Sahng-Kyoon Yoo and Dal-Ho Yoon
\prb {\bf 61},  11618, (2000) 
 \bibitem{c} 
 S.-Y. Lee, H. J. W. M\"{u}ller-Kirsten, D. K. Park, and F. Zimmerschied ; \prb {\bf 58}, 5554 (1998).

\bibitem{solo6}
Y.-B. Zhang, J.-Q. Liang, H.J.W. Muller-Kirsten, S.-P. Kou, X.-B. Wang and F.-C. Pu \prb {\bf 60}, 12886 (1999)
     
 \bibitem{ams}
J. M. Radcliffe, J. Phys. A: Gen. Phys {\bf 4} (1971), 313.

\bibitem{ams1}
John R. Klauder, \prd {\bf 19},  2349 (1979).

  
 
\bibitem{chud}
 E. M. Chudnovsky and L. Gunther, \prl {\bf  60}, 661 (1988)
   
\end{thebibliography}
\end{document}